\documentclass[conference]{IEEEtran}
\IEEEoverridecommandlockouts
\usepackage{cite}
\usepackage{amsmath,amssymb,amsfonts}
\usepackage{algorithmic}
\usepackage{graphicx}
\usepackage{textcomp}
\usepackage{ulem}
\usepackage{xcolor}
\def\BibTeX{{\rm B\kern-.05em{\sc i\kern-.025em b}\kern-.08em
    T\kern-.1667em\lower.7ex\hbox{E}\kern-.125emX}}
    \usepackage{multirow}
\DeclareMathOperator*{\argmax}{arg\,max}
\begin{document}

\title{A Validated Privacy-Utility Preserving Recommendation System with Local Differential Privacy
%\\ \thanks{This paper is supported by the chair Values and Policies of Personal Information, Institut Mines-Telecom, France, and European Union’s Horizon 2020 research and innovation programme under grant agreement No 830892, project SPARTA.}
}

\author{\IEEEauthorblockN{Seryne Rahali}
\IEEEauthorblockA{\textit{R3S Team/RST Department} \\
\textit{Samovar, Telecom SudParis,} \\ 
\textit{Institut Polytechnique de Paris} \\ 
Evry, France \\
Sirine.rahali@supcom.tn}
\and
\IEEEauthorblockN{Maryline Laurent}
\IEEEauthorblockA{\textit{Member of the Chair Values and }\\
\textit{Policies of Personal Information }\\
\textit{R3S Team/RST Department }\\
\textit{Samovar, Telecom SudParis, }\\
\textit{Institut Polytechnique de Paris }\\
Evry, France \\
ORCID: 0000-0002-7256-3721}
\and
\IEEEauthorblockN{Souha Masmoudi}
\IEEEauthorblockA{\textit{Member of the Chair Values and} \\
\textit{Policies of Personal Information} \\
\textit{R3S Team/RST Department} \\
\textit{Samovar, Telecom SudParis,} \\
\textit{Institut Polytechnique de Paris} \\
Evry, France \\
ORCID: 0000-0002-7194-8240}

\and
\IEEEauthorblockN{Charles Roux}
\IEEEauthorblockA{\textit{Telecom SudParis, Institut Polytechnique de Paris}\\
Evry, France \\
Charles.Roux@telecom-sudparis.eu}
\and
\IEEEauthorblockN{Brice Mazeau}
\IEEEauthorblockA{\textit{Telecom SudParis, Institut Polytechnique de Paris}\\
Evry, France \\
Brice.Mazeau@telecom-sudparis.eu}
}

\maketitle

\begin{abstract}
This paper proposes a new recommendation system preserving both privacy and utility. %, for individuals to get tailor-made services but not at the price of their privacy, and for recommenders to monetize the attention of consumers with targeted advertising, products and services selling.  
It relies on the local differential privacy (LDP) for the browsing user to transmit his noisy preference profile, as perturbed Bloom filters, to the service provider. 

The originality of the approach is multifold. 
First, as far as we know, the approach is the first one including at the user side two perturbation rounds - PRR (Permanent Randomized Response) and IRR (Instantaneous Randomized Response) - over a complete user profile.
Second, a full validation experimentation chain is set up, with a machine learning decoding algorithm based on neural network or XGBoost for decoding the perturbed Bloom filters and the clustering Kmeans tool for clustering users. 
Third, extensive experiments show that our method achieves good utility-privacy trade-off, i.e. a 90$\%$ clustering success rate, resp. 80.3$\%$ for a value of LDP $\epsilon = 0.8$, resp. $\epsilon = 2$. 
Fourth, an experimental and theoretical analysis gives concrete results on the resistance of our approach to the plausible deniability and resistance against averaging attacks. 
\end{abstract}

\begin{IEEEkeywords}
Local differential privacy, recommendation, privacy, RAPPOR, profiles perturbation, Bloom filters, neural networks, XGBoost, Kmeans
\end{IEEEkeywords}

\section{Introduction}
With the increase of online services, individuals are confronted with a lot of choices when making purchases. For better user experience, service providers rely on recommender systems helping users find items of interest. For this purpose, service providers are massively collecting and analyzing users’ data which may threaten users’ privacy. Hence, for individuals to get tailor-made services but not at the price of their privacy, as exposed in the survey of 2.000 people \cite{Poll2019}, and for recommenders to monetize the attention of consumers with targeted advertising, products and services selling, there is a strong need to elaborate new recommendation systems taking privacy and utility into account.

\textbf{Contributions}
This paper proposes a new recommendation system preserving both privacy and utility. The idea is to ensure protection against honest-but-curious entities - service providers and outsiders - while still getting useful recommendations. With that objective in mind, our approach relies on the LDP principle for the user preference profile to be perturbed at the browsing user side through a two-round processing - PRR (Permanent Randomized Response) and IRR (Instantaneous Randomized Response). That perturbation is an adaptation of the LDP-based RAPPOR approach \cite{Rappor} being made suitable for both classification and clustering tasks under local differential privacy.
As far as we know, this is the first time that a two-round perturbation processing has been applied to a complete user profile. 

With the objective of getting experimental utility vs privacy validation results, a full validation experimentation chain is set up with a recommender being implemented. At the recommender side, two successive mechanisms are performed: a machine learning decoding algorithm based on neural networks or XGBoost for decoding the perturbed Bloom filters and a clustering Kmeans tool for clustering users. It has to be noted that an appropriate user clustering leads straight to relevant recommendations for the user. 
The idea of the experiment is thus to assess how much users clustering is successful according to their perturbed preferences and a related privacy budget quantifying how much noise is included into their preferences. 

Our approach is validated through both extensive experiments and a security analysis. The experiments show that our method achieves a good utility-privacy trade-off with a 90$\%$ clustering success rate, resp. 80.3$\%$ for a value of LDP $\epsilon = 0.8$, resp. $\epsilon = 2$. The experimental and theoretical security analysis demonstrates that our approach supports both plausible deniability and resistance against averaging attacks. 

\textbf{Paper organization.}
Section~\ref{sec:Background} gives the useful background about Local Differential Privacy (LDP). Section~\ref{sec:RelatedWork} surveys existing LDP-related works highlighting their deficiencies. Then our approach is described in Sections~\ref{sec:SystemModel} and~\ref{sec:Phases}. Section~\ref{sec:SystemModel} introduces the system model with the actors, the utility metrics, the privacy properties and the threat model. Section~\ref{sec:Phases} details the step-by-step processing phases at both the user side and the recommender side.  
The three following sections provide a full evaluation of our approach. Section~\ref{sec:Performance} studies the utility performance achievements of our decoding and clustering algorithms according to several experimental conditions (privacy budget, Bloom filter parameters). Section~\ref{sec:Security} gives a security analysis of our scheme with regard to the plausible deniability and averaging attacks.   
Section~\ref{sec:Trade-off} discusses the utility vs privacy trade-off. Conclusions are given in Section~\ref{sec:Conclusion}.

\section{Local Differential Privacy Background}
\label{sec:Background}
Local Differential Privacy (LDP) has its roots in Differential Privacy (DP) works \cite{DPP,dp}. DP matches the global model where a trusted third party uses DP to produce statistics over a dataset, while withholding information about individuals in the dataset. LDP \cite{Ldp} matches the local model where data can be perturbed right at the source, locally to the user, thus leading to higher privacy guarantees as the trusted third party is no longer necessary. 

%Unlikely to Differential Privacy (DP) - designed for producing statistics over a dataset while withholding information about individuals in the dataset - Local Differential Privacy applies on the  enables the users to perturb their data locally

\textbf{Definition.} 
A randomized algorithm $M$ satisfies the $\epsilon$ local differential privacy \cite{Ldp} where ${\epsilon > 0}$ if for all pairs of the client 's values x and y and for all $S \subseteq$ range($M$):
  \begin{equation}
      \Pr[M(x)\in S] \leq e^{\epsilon}\Pr[M(y)\in S]
  \end{equation}

The definition introduces $\epsilon$, known as the privacy budget or the privacy loss. It controls to what extent the output of an algorithm depends on the input, and thus it reflects the desired level of privacy. The smaller is the privacy budget $\epsilon$, $\epsilon$ being a positive value, the higher is the privacy level. 

%\subsubsection{Randomized response}
% Even though the notion of local differential privacy has gained interest only in recent years, the idea behind is much older. It came mainly from the randomized response works ~\cite{randomized}  which is an old surveying technique introduced in 1965 by Warner. It enables collecting  information about sensitive topics while preserving users privacy. For instance, given the question " Are you diagnosed with VIH" the respondent flips a coin. If the coins comes up heads then the respondent answers truly to the question. In the other case where the coin comes up tails, he answers randomly with yes or no. This way, we guarantee plausible deniability for the user as the receiver can not distinguish the true answer from the random one. Yet, if the same query is repeated for the same user many times the privacy level would degrade. Another limit to this technique is being suitable only for binary questions.

 \section{Related work}
\label{sec:RelatedWork}
This section makes an overview first on some LDP well-known use cases, and then on some privacy preserving systems including LDP recommender systems.
 
 \textbf{LDP use cases.} LDP concepts are applied in various domains in the literature. For instance, Google proposed RAPPOR \cite{Rappor} so as to collect statistics about malicious URLS from users browsers without causing privacy threats. The proposed solution relies on encoding the values to be sent, by applying two $\epsilon$-LDP perturbation levels. In Microsoft, LDP enables collecting data about the time spent by users in different applications, therefore, identifying its favorite ones and improving the users' experience \cite{Microsoft}, while still preserving privacy. For reducing possible privacy leakage in deep learning models, \cite{latent} proposes LATENT as an intermediate layer designed to satisfy LDP in deep learning models. The suggested solution enables a data owner to perturb data at the owner's device before the data reach out an untrusted machine learning service. 

 \textbf{Privacy preserving recommender systems.} The need for privacy preservation in recommender systems triggered research efforts in the last decades. Two main axes based on cryptography and data perturbation are investigated. Kaaniche et al. \cite{CoWSA2020} designed a privacy-preserving framework for recommender systems. They suggested that a user perturbs his profile relying on a collaborative secure computation, that incorporates intermediate nodes between end-users and service providers. Their framework generates additional computation overhead. In \cite{can,can2}, Canny proposes cryptographic protocols to preserve privacy in recommender systems. The suggested scheme uses matrix projection and factor analysis. Both techniques result in heavy computational and communication overhead. Aïmeur et al. \cite{cryp} designed the Alambic system where users private data are shared between the service provider and a semi-trusted third party. The whole public key infrastructure is adopted so as to ensure data protection. Yet, only if the service provider doesn't collude with the semi-trusted third party, the user privacy is protected. Recently, Kim et al. \cite{spirel} suggested SPIREL, a location based recommender system under LDP. The framework uses the Optimized Randomized Response \cite{Ldp} approach to perturb the transition patterns and refers to the piecewise mechanism \cite{piece} in order to perturb gradients. Both suggested algorithms are $\epsilon-LDP$. The solution drawback is its high communication cost, as the transition patterns are encoded in a $n^2$-sized bit-array ($n$ denoting the domain of possible locations) by the user before perturbing them with the Optimized Randomized Response. Moreover, the framework takes into account only one location transition from the user's check-in history. Nevertheless, using LDP to protect only one transition doesn't fit well in practice as the user may have many transitions to report. The extension of SPIREL to report many items under LDP is far from obvious as the process relies on the gradient perturbation. While most of the existing works tackle the issue of only one item perturbation, BLIP \cite{Blip} focuses on perturbing a whole profile using the LDP mechanism. However, the achieved utility in terms of recall does not exceed $0.26$ for $\epsilon$ = 3. Moreover, the utility computation is performed on perturbed Bloom filters, and does not refer to any decoding algorithm, which is not compatible with an integration into a recommender system. 
 Furthermore, as the similarity computation between profiles is done at user side each time a new node joins the system, the solution can be considered as cost-prohibitive. 
 
\section{System Model and Overview}\label{sec:SystemModel}
This section details our full system model, and an overview of our approach. It presents the actors, the metrics for measuring utility, the supported privacy properties along with the considered threat model. 

\subsection{Actors}\label{sec:network}
Our system model is composed of the following entities: 
%We dispose of a web search engine along with an advertisement and recommendations provider. Users trust neither the web search engine nor the recommendations provider. Consequently, they opt to perturb their preferences locally. We distinguish then the following entities.
\begin{itemize}
    \item Client $C$: a browsing user submits a query to a Web Search Engine $WSE$ along with his profile made of individual attributes (preferences). $C$ who cares about his privacy, sends to $WSE$ his perturbed preferences according to the LDP method and a privacy budget $\epsilon$. 
    \item Web Search Engine $WSE$: the $WSE$ server responds to the client’s query according to the transmitted preference profile. $WSE$ can suggest its own recommendations to $C$, and can take on the role of a proxy between $C$ and third parties $TP$ by relaying the profile and the $TP$'s recommendations back to $C$. Note that the intermediate $WSE$ between $C$ and $TP$ could be any Service Providers. 
    \item Third Party $TP$: $TP$ is an advertising provider or any recommendation provider. Its main goal is to provide users with targeted recommendations. 
\end{itemize}

\subsection{An overview}
\label{sec:Overview}
\begin{figure*}[!ht]
\begin{center}%	\centering
		\includegraphics[width=0.86\textwidth]{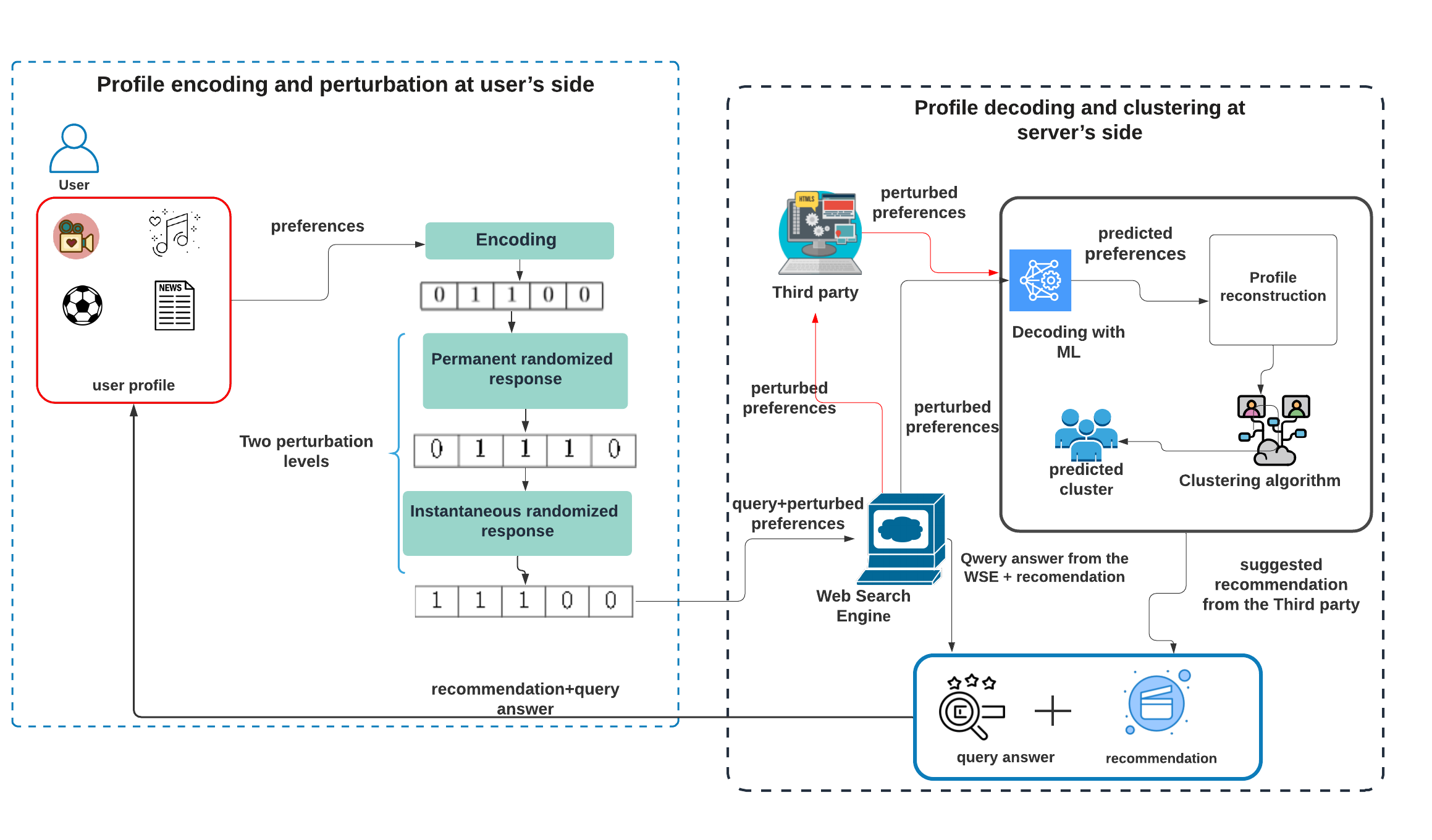}
	\caption{System overview including a user $C$ and a server ($TP$ or $WSE$)}
	\label{overview-Fig}
\end{center}
\end{figure*}

To avoid leaking his preferences while still getting appropriate recommendations, a user $C$ can decide to send a search query to $WSE$ along with his perturbed preferences. The full processing for perturbing the preferences is detailed in subsection \ref{sec:perturbation} and includes the following steps which are depicted in Figure \ref{overview-Fig}: 

%Every user $C$ has his profile stored locally on his browser. The latter contains the attributes describing $C$'s personal preferences for music, movies, sports, etc.
%Whenever $C$ issues a query to the web search engine (WSE), he transmits his preferences as well in order to get appropriate recommendations. Being aware of the threat caused by revealing his personal preferences, the client proceeds to:

\begin{enumerate}
    \item $C$ preferences using a Bloom filter \cite{Rappor}
    \item Execute the permanent randomized response (PRR) step over the Bloom filter \cite{Rappor}
    \item Execute the Instantaneous Randomized Response (IRR), as a second perturbation step \cite{Rappor}, over the modified Bloom filter 
    \item Send the query along the perturbed preferences to $WSE$ 
\end{enumerate}

Upon receiving the request, $WSE$ forwards the perturbed profile to the recommender, either $TP$ or $WSE$, and next referred for clarity as $TP$. $TP$ next needs to decode the given preferences using a machine learning algorithm, resulting in a reconstructed approximate user profile (cf. subsection \ref{sec:decoding-subs}). Then based on the obtained profile and the similarity among profiles, $TP$ executes a clustering algorithm (cf. subsection \ref{sec:clustering-subs}) and classifies $C$ in one of the group of users, for next delivering targeted recommendations suitable for that group of users to $C$ via $WSE$.
    
\subsection{Metrics for Measuring Utility}\label{sec:functional}
   Our recommendation system should preserve the utility - i.e. the adequacy between $C$'s expectations and the returned recommendations - for maintaining the user experience. This means that the user clustering being performed by recommenders $WSE$ or $TP$ on the perturbed $C$ profile should be as close as possible to the one achieved with the unperturbed $C$ profile. 
   
   To measure the utility, the metrics well known for clustering algorithms - accuracy, precision, recall, and F1 score - rely on the following values: 

    \begin{itemize}
    \item \textbf{True Positive (TP).} The instance belongs to the class, and is predicted to be in the class.
    \item \textbf{False Positive (FP).} The instance does not belong to the class, but it is predicted as a class member.
    \item \textbf{False Negative (FN).} The instance belongs to a class, but it is predicted as not being a member of that class.
    \end{itemize}

    The following metrics are used for measuring utility: 
\begin{enumerate}
    \item \textbf{Accuracy}. The rate of the correct predictions out of all the predictions. Accuracy is sensitive to class imbalance, and is expressed as follows:
    \begin{equation}
       \text{Accuracy} = \frac{\text{Number of correct predictions}}{\text{Total number of predictions}}
    \end{equation}

    \item \textbf{Precision}. The proportion of items correctly identified to be within a class $i$ out of all the items identified to belong to that class. 
    A low precision helps to identify where the model made incorrect classification. 
    It is defined as:   
    \begin{equation}
       \text{Precision}_{i} = \frac{\sum TP_{i}}{\sum TP_{i}+FP_{i}}
    \end{equation}   

%A higher precision indicates that the classifier is accurate.

   \item \textbf{Recall.} The proportion of items that should have been annotated with a given label, and that were actually annotated with that label. This is a measure of completeness and is formally defined as follows:
   \begin{equation}
      \text{Recall}_{i} = \frac{\sum TP_{i}}{ \sum TP_{i}+FN_{i}}  
   \end{equation}

   \item \textbf{F1 \-score.} A weighted average of recall and precision. Thereby, the metric takes into account both false positives and false negatives, as follows:
   \begin{equation}
     \text{F1} = 2*\frac{Recall * Precision}{Recall + Precision}
   \end{equation}
\end{enumerate}

  %Accurate clustering assignment:} grouping users according to their perturbed profiles should not be very different from their clustering before profile perturbation. In other words, the perturbation should not affect too much the clustering quality, in order to maintain the utility through a good user experience. 

\subsection{Privacy Properties}\label{sec:security} 
This section defines the basic privacy properties supported by our approach. 
    \begin{itemize}
        \item \textbf{Plausible deniability.} This property is granted through the use of an algorithm that satisfies the LDP definition as an obfuscation scheme. In fact, by observing the output of the algorithm, an adversary can not tell with high confidence whether a particular preference was definitely used as an input of the algorithm.
        \item \textbf{Resistance against averaging attacks.} Any adversary having knowledge of several versions of the perturbed Bloom filters can not find out the original preferences through an averaging attack. 
    \end{itemize}

%\subsection{Performance properties}\label{performance}
%     Under the above mentioned privacy requirements, our system should also satisfy some performance conditions.
%     \begin{itemize}
%    \item \textbf{Scalable communication overhead.} The communication overhead should be acceptable. Consequently, the response time should remain sustainable despite the number of transmitted attributes.
%    \item \textbf{User friendly.}  In the proposed solution, the user is only asked to provide a value for the desired level of privacy. As a result, he would get satisfactory recommendations without getting too implicated in the complexity of the system. 
%    \item \textbf{Low computation cost.} Tasks requiring high computation costs are outsourced to other parties. Therefore, we leave a minimal computational burden to the client.
    %Tasks performed at the client side aren't greedy in terms of computation.
%    \end{itemize}

\subsection{Threat Model and Privacy Games}\label{sec:threat}
Our threat model considers a honest-but-curious adversary attempting to learn the preferences of client $C$ from the perturbed Bloom filters sent by client $C$. We next define three privacy games, with regard to two adversaries provided with the following abilities: 
  \begin{enumerate}
      \item \textbf{Basic Adversary $BA$.} $BA$ is not provided with any Machine Learning algorithms. $BA$ is an outsider, i.e. external to our system.
      \item \textbf{Advanced Adversary $AA$.} $AA$ is provided with Machine Learning algorithms. $AA$ is an honest-but-curious WSE or an honest-but-curious $TP$.
  \end{enumerate}

\subsubsection{The Plausible Deniability Game Conducted over one Preference by a Basic Adversary} \label{sec:onep}
An outsider $BA$ is challenged with the following privacy game: 
\begin{itemize}
\item \textbf{Setup.} The challenger $C$ provides the adversary $BA$ with a set of $N$ preferences, the Bloom filter size $M$, $k$ hash functions and the privacy budget $\epsilon$. 
\item \textbf{Training phase.} Upon receiving the mentioned parameters, $BA$ is given a polynomial computation time in order to compute some Bloom filters for the transmitted $N$ preferences. Therefore, he can reconstruct a database $DBF$ of perturbed Bloom filters.
\item \textbf{Challenge phase.} $C$ selects a preference $i$ and sends to $C$ a related perturbed Bloom filter. $BA$ sends back to $C$ his guess about the preference.
\end{itemize}

$BA$ wins the game if he is able to correctly guess the encoded preference.

If the probability to win the game is negligible, then our approach is said to be resistant against the plausible deniability attack conducted over one preference by a basic adversary.

\subsubsection{The Plausible Deniability Game Conducted over multiple preferences by the Advanced Adversary} \label{sec:multi}
The insider $AA$ is challenged with the following privacy game: 
\begin{itemize}
    \item \textbf{Setup.} The challenger $C$ provides $AA$ with a set of perturbed preferences along with their original values. 
    \item \textbf{Training phase.} $AA$ trains his ML algorithm with provided values.
    \item \textbf{Challenge phase.} $C$ samples a set of preferences from his dataset, and sends to $C$ a related perturbed Bloom filter. $AA$ sends back to $C$ his guess about the set of preferences, which comes down to correctly decoding the perturbed Bloom filters.
\end{itemize}

The adversary wins the game if he gets high precision and recall.

If the probability to win the game is negligible, then our approach is said to be resistant against the plausible deniability attack conducted over multiple preferences by an advanced adversary.

\subsubsection{Averaging Game} \label{sec:averaging-att}
Any $BA$ or $AA$ adversary is challenged to find out the original preferences from a set of perturbed Bloom filters issued over the same set of $C$'s preferences.

\section{Detailed Processing Phases of our Approach}
\label{sec:Phases}

In this section, we detail the main three phases of the new recommendation system, an overview of which is described in Section \ref{sec:Overview}.

%Every user $C$ has his profile stored locally on his browser. The latter contains the attributes describing $C$'s personal preferences for music, movies, sports, etc.
%Whenever $C$ issues a query to the web search engine (WSE), he transmits his preferences as well in order to get appropriate recommendations. Being aware of the threat caused by revealing his personal preferences, the client proceeds to:

\subsection{Perturbation Phase at the Client}\label{sec:perturbation}
The perturbation is handled at $C$'s device with the objective to obfuscate $C$'s preferences. It has to be noted that the set of preferences forms a profile composed of one or many categories where each category is denoted by $C_{i}$, $i$ $\in \{1...l\}$ where $l$ denotes the maximum number of supported categories. Each category is composed of groups of interests I$_{ij}$ (i.e. $j \in \{1...g\}$) where $g$ represents the maximum number of groups of interests for category $i$.

\begin{enumerate}
     \item \textbf{Encoding}. As the first step of the perturbation process, the encoding algorithm maps $C$’ preferences into bits in a Bloom filter. 

The first operation is to compute the optimal Bloom filter size $m$, given the maximum number of preferences encoded into the Bloom filter (i.e. $n = l * g$ if each category includes the same number of groups of interests), and a fixed false positive rate $f_{p}$, according to equation \ref{equa1} \cite{bloom}: 

%Indeed, given the number of preferences to report $n$ and a fixed probability rate $f_{p}$ the algorithm computes the optimal size \cite{optimals} of the Bloom filter $m$ according to the equation \ref{equa1} \cite{bloom}.

\begin{equation} \label{equa1}
    m =- \frac{n\times\ln({f_{p}})}{(\ln{2})^{2}}
\end{equation} 

Then, the optimal number of hash functions \cite{optimals} is computed as given by equation \ref{right}.

\begin{equation}\label{right}
    k=\frac{m}{n}\times\ln{2} 
\end{equation}

\begin{figure}
\begin{center}
\includegraphics[scale=0.50]{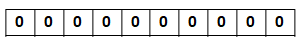}
\caption{Empty Bloom filter, $m=10$}
\label{empty}
\end{center}
\end{figure} 

After selection of $m$ and $k$ appropriate values, the Bloom filter $B$ is initialized with all "0" values, as given in Figure \ref{empty}. For feeding $B$ with the set of preference values $v$, $C$ first applies the $k$ hash functions to $v$, and feeds $B$ with the hash output providing indices. For illustration, let us consider a Bloom filter of size $m=10$ bits, with $k=2$ hash functions ($h1, h2$), and two preferences $\{$Action, Fantasy$\}$ to be included into the Bloom filter. As given in Figure \ref{action}, $C$ first computes the two hashes of the preference $Action$, and gets the results $h_{1}$(Action)$=3$ and $h_{2}$(Action)$=6$, thus leading to positioning the $3^{rd}$ and the $6^{th}$ bits of the Bloom filter to value $1$. The same applies for the preference $Fantasy$ as depicted in Figure \ref{fantasy} where $h_{1}$(Fantasy)$=2$ and $h_{2}$(Fantasy)$=8$.\\

%Then, we apply the two hash functions to get the appropriate indices. Accordingly, for the word "Action", we get $h_{1}$(Action)$=3$ and $h_{2}$(Action)$=6$. Thereby, we should set the $3^{rd}$ and the $6^{th}$ bits in the Bloom filter equal to 1 to represent the word action as demonstrated by figure \ref{fantasy}.
%As for the word "Fantasy", we get that $h_{1}$(Fantasy)$=2$ and $h_{2}$(Fantasy)$=8$. The result of inserting the word Fantasy into the Bloom filter is demonstrated in figure \ref{fantasy}.

\begin{figure}%[H]
\begin{center}
\includegraphics[scale=0.50]{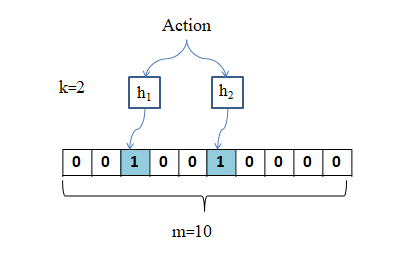}
\caption{Feeding the Bloom filter with the preference "Action", $m=10$, $k=2$ }
\label{action}
\end{center}
\end{figure} 

\begin{figure}%[H]
\begin{center}
\includegraphics[scale=0.50]{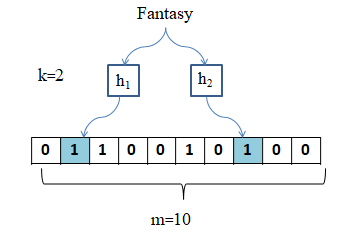}
\caption{Feeding the Bloom filter with the preference "Fantasy", $m=10$, $k=2$}
\label{fantasy}
\end{center}
\end{figure}

\item \textbf{Permanent Randomized Response (PRR).} This first level perturbation applies over the Bloom Filter $B$ obtained through the encoding phase. This step is executed once over a set of preferences $v$, whatever the number of search requests done by $C$ to $WSE$. A noisy bit is derived from each bit of $B$ thus resulting in a perturbed Bloom filter vector $B^{'}$. The derivation is compliant with the RAPPOR works \cite{Rappor} and considers the following probabilistic processing: 

\begin{equation}\label{pro}
{ B^{'}[i] = }
\begin{cases}
1 & \text{with probability $ \frac{1}{2}f$}\\
0 & \text{with probability $ \frac{1}{2}f$}\\
B[i]& \text{with probability $1-f$}\\
\end{cases}
\end{equation}
Where $f$ is the privacy level parameter configured by $C$.

The obtained bit vector $B^{'}$ remains stored and known to $C$ only. 
This first level perturbation PRR algorithm is $\epsilon$-differential privacy with the following quantified $\epsilon_{1}$ privacy budget: 
\begin{equation}
    \epsilon_{1}=k \ln{\left( \frac{1-\frac{1}{2}f}{\frac{1}{2}f}\right)}
\end{equation}\\

\item \textbf{Instantaneous Randomized Response (IRR).} To guarantee stronger short-term privacy, this second level perturbation \cite{Rappor} is executed for each request done by $C$ to $WSE$. After getting $B^{'}$, the user initializes a bit vector $S$ with all zeros and then applies the following probabilistic processing: 

\begin{equation}\label{IRR}
   p(S[i]=1)
 \begin{cases}
     q & \text{if $ B^{'}[i]=1$ }\\
     p & \text{if $ B^{'}[i]=0$}\\
 \end{cases} 
\end{equation}
Where $p$ denotes the probability of flipping a bit that equals to 0 into 1 whereas $q$ represents the probability of keeping bits equal to 1.

This second level perturbation IRR algorithm is $\epsilon$-differential privacy with the following quantified $\epsilon_{2}$ privacy budget \cite{Rappor}: 
\begin{equation}
 \epsilon_{2}=k \ln{\left( \frac{q^{'}(1-p^{'})}{p^{'}(1-q^{'})}\right)}
\end{equation}

Where $p^{'}$, resp. $q^{'}$, is the probability of observing 1 given that the same Bloom filter bit was set to 0, resp. 1, as defined in the following equations. 
\begin{equation}
    p^{'}= \frac{1}{2}f q+(1-\frac{1}{2}f) p 
\end{equation}
\begin{equation}
    q^{'}= (1-\frac{1}{2}f)(1-q)+\frac{1}{2}f (1-p)
\end{equation}
\end{enumerate}

\subsection{Decoding Phase by the Recommender}\label{sec:decoding-subs}
Upon receiving $C$'s perturbed preferences, $TP$ decodes the received $C$'s perturbed preferences into some approximate preferences, based on a machine learning algorithm. The problem is modeled as a multiclass classification task, aiming at  predicting the classes of perturbed Bloom filters. 
For instance, given the music category, which contains eight classes (groups of interest): classical, jazz, pop... $TP$ should identify for each perturbed Bloom filter its right label. Two machine learning algorithms - neural network and XGBoost - were selected for their ability to work on perturbed data. Both algorithms are calibrated to fit the specificities of the two following datasets, thus resulting into 2 configurations as  detailed below:

%This fact stems in a difficulty of reconstructing the original preferences without referring to a machine learning algorithm. Thus, we model the problem as a multiclass classification task.
%For instance, given the category music, which contains eight classes: classical, jazz, pop... the server should identify for each perturbed Bloom filter its right label. We selected neural network and XGBoost as the best algorithms to fit our choice. This is due to their ability to work on perturbed data.

\begin{itemize}
    \item \textbf{Preference dataset.} The set of preferences includes 3 categories - movies, music and sports - and 7 classes for movies, 8 for music and 12 for sports.
    \item \textbf{Flight dataset.} The set of preferences includes 3 categories - destination and flight class type - which are made up 11 classes for destination and 3 for flight class type.
\end{itemize}

%\begin{table}
%\caption{The preference dataset } 
%\begin{center}
%\centering
%   \begin{tabular}{|c|c|}
%   \hline
%      Category & Number of classes  \\ 
%     \hline  Music & 8   \\
%    \hline   Movies & 7 \\
%    \hline Sports& 12\\ 
%    \hline
%\end{tabular}
% \label{a}
%\end{center}
%\end{table}

%\begin{table}
%\caption{The flight dataset } 
   % \centering
%\begin{center}
%   \begin{tabular}{|c|c|}
%        \hline
%       Category & Number of classes  \\ \hline
%     Destination & 11 \\ \hline
%     Flight class type & 3 \\ \hline
%\end{tabular}
% \label{tab:b}
%\end{center}
% \end{table}

\begin{enumerate}
 \item \textbf{Neural network configuration.}
     The neural network is composed of an input layer which is fed with the perturbed preferences, two hidden layers, and an output layer. Two dropouts are introduced to mitigate the overfitting issue. For both of our datasets, the layers of the algorithm are configured according to the parameters given in Table \ref{net}.

%\begin{figure}[H]
%\begin{center}
%\includegraphics[scale=0.3]{model(1).png}
%\caption{The different layers of our neural network model}
%\label{model}
%\end{center}
%\end{figure} 

\begin{table}
%\captionsetup{justification=centering}
\caption{Neural network configuration} 
\begin{center}%\centering
\begin{tabular}{|c| c|}
\hline                       
Parameter& {Value}\\  
\hline
\multirow{3}{*}{The number of nodes}
& Input layer: 80\\
&First hidden layer: 60\\
&Second hidden layer: 50\\

\hline           
Loss function& Categorical crossentropy
\\
\hline
\multirow{2}{*}{The activation function}
& Output layer: Softmax\\
& Other layers: ReLu\\
\hline
Number of epochs& 25
\\
\hline
Batch size& 70
\\
\hline
Optimizer& Adam
\\
\hline
\end{tabular}
\label{net}
\end{center}
\end{table}

\item \textbf{XGBoost configuration.}
XGBoost is a gradient boosting algorithm. Table \ref{xgb} gives the parameters calibrated for each dataset to optimize the model's performances. As can be shown, the configuration is slightly the same, except for parameter Subsample. 

\begin{table}
%\captionsetup{justification=centering}
\caption{XGBoost configuration} 
\begin{center}
%\centering
\begin{tabular}{|c|c|c|}
   \hline
	Parameter & Preference dataset& Flight dataset\\ \hline
	N${\_}$estimators &100&100 \\
	\hline
	Max${\_}$depth&3&3 \\
	\hline
	Min${\_}$child${\_}$weight &1&1 \\
	\hline
	Subsample &$0,8$&$0,9$ \\
	\hline
	Gamma &$0,4$&$0,4$\\
	\hline
\end{tabular}
 \label{xgb}
\end{center}
\end{table}
\end{enumerate}

\subsection{Clustering Tool and Accuracy Measurement}     \label{sec:clustering-subs}

Clustering is done with Kmeans, considering $K=4$ clusters for a number of profiles that is equal to 80.000 and $K=5$ clusters for 90.000 profiles (used later to study the privacy utility trade-off).
This choice of $K$ is based on the elbow method \cite{elbow}. The main idea behind the technique is to run the Kmeans algorithm for different values of clusters and to calculate the Within Cluster Sum of Squares (WCSS). According to this metric, the variability of observations is computed in each cluster. A cluster that has a low value of WCSS is more homogeneous than a cluster with a high WCSS value. 
Formally, the objective is to minimize the following function. 
\begin{equation}
    WCSS= \sum_{i \in n}(X_{i}-Y_{i})^2
\end{equation}

\begin{figure}
\begin{center}
\includegraphics[scale=0.65]{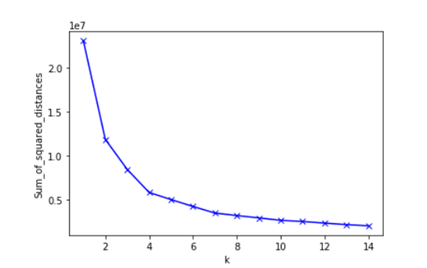}
\caption{Searching for the optimal number of clusters}
\label{K}
\end{center}
\end{figure}

Where $Y_{i}$ represents the centroid for observation $X_{i}$.
Then, for a range of $k$ numbers, the WCSS variation is plotted with respect to $k$. The optimal value of $k=4$ was obtained through an experiment we did over 80.000 cleaned traces from a Qwant dataset to the Kmeans, and a range of $k$ varying between 1 and 15. As depicted in Figure \ref{K}, there is a sudden huge drop in the WCSS value when increasing the number of clusters from 1 to 2, and a second drop - not as huge as the first one - at the cluster number 4. As the WCSS maintains a minimum value starting from $k=4$, we deduce an optimal value of $k = 4$.

Unlikely to usual recommendation systems, our LDP-recommendation system works on perturbed profiles instead of the true users profiles. It is therefore necessary to adapt the accuracy measurement for evaluating the ability of the algorithm to group same-profile users into the same cluster. 
In our case, in a first experience, Kmeans is  applied over a set of original (unperturbed) profiles, thus resulting in some cluster labels, as classically done. In a second experience, Kmeans is fed with profiles which have been first perturbed and then decoded. The clustering results are recorded and then used in a final step for comparing the clustering results with/without perturbation and for measuring accuracy. The more matches we get, the higher our accuracy.

\section{Performance Analysis of the Decoding and Clustering Algorithms}
\label{sec:Performance}
%This section evaluates the performances and the utility preservation for both of our decoding and clustering algorithms, according to several parameters, e.g. privacy budget, Bloom filter size. 

\subsection{Decoding Algorithms Evaluation}
As shown in Table \ref{tab:comprelated}, neural network outperforms XGBoost for both datasets with regard to the measured accuracy (for $K=4$ clusters). This result is confirmed in Table \ref{tab:comprelated} for low perturbation level ($\epsilon = 2$) as well as high perturbation level ($\epsilon = 0.80$). It gives higher precision, recall and f1-score and thus higher clustering success rates. 
%We firstly compare the performances of both XGBoost and neural  network in terms of accuracy. As demonstrated by figure \ref{acc}, neural network outperforms XGBoost in both dataset in the measured accuracy.
\begin{figure}
%\centering
\begin{center}
\includegraphics[scale=0.50]{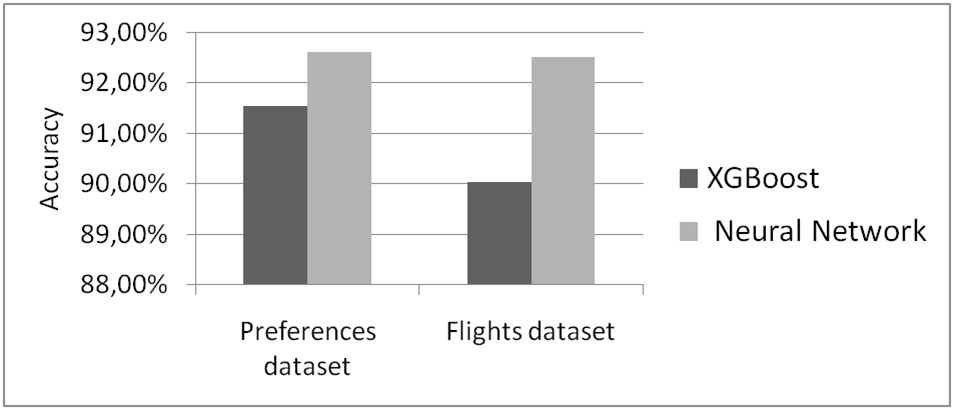}
\caption{Accuracy of XGBoost and neural network, $\epsilon = 0.8$, $K=4$}
\label{acc}
\end{center}
\end{figure} 

%We evaluate also the algorithms according to their classification results in both low perturbation level ($\epsilon = 2$) and high perturbation ($\epsilon = 0.80$). Indeed, when $\epsilon$ was set to be high, indicating that the perturbation is low, neural network has better precision, recall and fscore in all the classes than XGBoost as demonstrated in figures \ref{R}.(a) and \ref{R}.(c).

%Even when we increased the perturbation level by setting $\epsilon=0.80$, as depicted in figure \ref{R}.(d) neural network  was able to decode correctly and with higher classification results the majority of the classes while XGBoost has lower precision and recall as demonstrate in figure \ref{R}.(b)
%Nevertheless, we remark that both algorithms exhibit good results in overall.

\begin{table*}
\begin{center}
\caption{Decoding the movies category with XGBoost and neural networks. Validation set=40.000, $fp=0.10$, $K=4$}
\resizebox{\textwidth}{!}{
\begin{tabular}{c|c|c|c|c|c|c|c|c|c|c|c|c|c|c}
\hline
 Metrics & \multicolumn{4}{c|}{ Precision } & \multicolumn{4}{c|}{Recall} & \multicolumn{4}{c|}{F1-score} & \multicolumn{2}{c}{Support} \\ \hline
   & \multicolumn{2}{c|}{ Neural Network } & \multicolumn{2}{c|}{XGBoost} & \multicolumn{2}{c|}{ Neural Network } & \multicolumn{2}{c|}{XGBoost} & \multicolumn{2}{c|}{ Neural Network } & \multicolumn{2}{c|}{XGBoost} &  Neural Network  & XGBoost \\ \hline
& $\epsilon = 0.80$ & $\epsilon = 2 $ & $\epsilon = 0.80$ & $\epsilon = 2 $ & $\epsilon = 0.80$ & $\epsilon = 2 $ & $\epsilon = 0.80$ & $\epsilon = 2 $ & $\epsilon = 0.80$ & $\epsilon = 2 $ & $\epsilon = 0.80$ & $\epsilon = 2 $ & $\epsilon = 0.80$ and $\epsilon = 2 $ & $\epsilon = 0.80$ and $\epsilon = 2 $ \\ \hline \hline
Action & 0.95 & 0.94 & 0.82 & 0.90 & 0.93 & 0.99 & 0.81 & 0.88 & 0.94 & 0.96 & 0.81 & 0.89 & 8550  & 5751  \\ \hline
Comedy & 0.92 & 0.96 & 0.81 & 0.86 & 0.95 & 0.96 & 0.81 & 0.86 & 0.93 & 0.96 & 0.81 & 0.86 & 5693 & 5693 \\ \hline
Drama & 0.87 & 0.92 & 0.80 & 0.89 & 0.81 & 0.92 & 0.92 & 0.93 & 0.84 & 0.92  & 0.86 & 0.91 & 5691 & 8550  \\ \hline
Fantasy & 0.92 & 0.95 & 0.83 & 0.85 & 0.91 & 0.96 & 0.79 & 0.85 & 0.92 & 0.95 & 0.81 & 0.85 &  5751 & 5690 \\ \hline
Horror & 0.82 & 0.89 & 0.82 & 0.83 & 0.88 & 0.97 & 0.80 & 0.86 & 0.85 & 0.93 & 0.81 & 0.84 & 5721 & 5691 \\ \hline
Romance & 0.90 & 0.98 & 0.82 & 0.86 & 0.93 & 0.90 & 0.80 & 0.85 & 0.91 & 0.94 & 0.81 & 0.86 &  5690 & 5721  \\ \hline
Thriller & 0.94 & 0.96 & 0.89 & 0.87 & 0.91 & 0.81 & 0.71 & 0.74 & 0.92 & 0.88 & 0.79 & 0.80 & 2904 & 2904 \\ \hline

  \end{tabular}}
 \label{tab:comprelated}
\end{center}

\end{table*}

\subsection{Clustering Evaluation}
\begin{figure}
\begin{center}
%\centering 
\text{[Bloom filter size for $\epsilon=0.85$, $f_{p}=0.1$ and $k=3$]\label{d} }
{\includegraphics[scale=0.75]{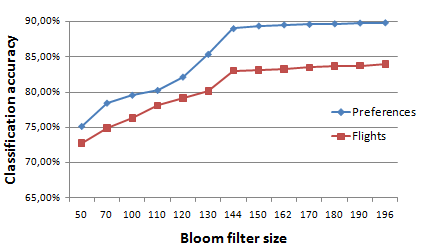}}
\text{[Hash functions for $M=144$, $\epsilon=0.85$ and $f_{p}=0.1$]\label{b}}
{\includegraphics[scale=0.78]{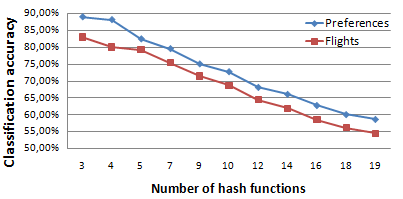}}
\text{[Privacy budget, $M=144$, $f_{p}=0.1$, k=$4$]\label{V}}
{\includegraphics[scale=0.70]{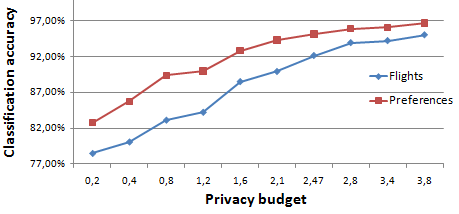}}
\caption{Impact of several parameters on the clustering results, $K=4$}
\label{Impact}
\end{center}
\end{figure}

This subsection analyses the influence of different parameters on the clustering results, including the privacy budget value $\epsilon$, the Bloom filter size $M$ and the number of hash functions $k$.
\begin{itemize}
    \item \textbf{Bloom filter size.} 
    As expected in Figure \ref{Impact}, larger $M$ results in higher classification accuracy, as the Bloom filter collision rate decreases. Yet, this comes at a cost in memory since increasing $M$ leads to larger Bloom filter sizes. As shown, the benefit for high values of $M$ diminishes when $M$ exceeds 144 for this experimental setup.    This observation is due to the hash collisions starting to vanish. Next observations are thus considering $M=144$.  

    \item \textbf{Number of hash functions.} Our experiment considers a minimum value of hash functions of 3, which corresponds to the optimal number of hash functions for $M=144$, according to the equation \ref{right}. As depicted in the figure, the classification accuracy decreases when the number of hash function increases. This stems from an increasing number of hash collisions.

    \item \textbf{Privacy budget.}
    As expected in Figure \ref{Impact}, the clustering accuracy is an increasing function of the privacy budget. Indeed, the higher the privacy budget, the lower the perturbation level, and the higher the accuracy. The preference dataset achieves better clustering results. This might be due to the dataset characteristics, the number of categories, the number of classes per category...
\end{itemize}
    
\section{Security analysis}
\label{sec:Security}
This section is dedicated to the security analysis with regard to the threat model defined in subsection \ref{sec:threat}. 

\subsection{Plausible Deniability over one Preference}
Referring to the game described in Section \ref{sec:threat}, given the perturbed Bloom filter $y$ sent by the challenger $C$ and the preference universe $D$, the strategy for the basic adversary $BA$ is to compute:

%We provide a theoretical proof for this game. In fact, given the Bloom filter $y$ that the challenger has sent, the optimal attack strategy for $A$ would be the one such as:

\begin{equation}
\begin{split} 
\text{guess(y)} & = \argmax_{v\in D} Pr[v|y] \\
   & =  \argmax_{v\in D} \frac{\pi({{v})}. Pr [{\phi_{RAPPOR}(v)=y)}]}{\sum_{a \in D} \pi{(a)}Pr[{\phi_{RAPPOR}(a)=y}]}
\end{split}
\end{equation}

 Where $\phi_{RAPPOR}$ denotes the perturbation mechanism, $D$ is the universe of the preferences to enter into a Bloom filter, $ \pi({v})$ is  the prior probability of ${v}$ and is equal to $\frac{1}{|D|}$ for all $v \in D$.
 
%If  guess$(y)$ is  high then the adversary can more easily predict the true input.
For computing for each value $v$ the probability that the perturbation mechanism outputs $y$, $BA$ can refer to the following probability expressed in Equation \ref{dec} \cite{delta}.

\begin{equation}\label{dec}
Pr \left[ B^{'}[i]= \phi_{RAPPOR}(B[i])=1\right]= 
\begin{cases}
\frac{e^{\frac{\epsilon}{2 \Delta}}}{e^{\frac{\epsilon}{2 \Delta}}+1} &\text{if $B[i]=1$} \\
\frac{1}{e^{\frac{\epsilon}{2 \Delta}}+1} &\text{if $B[i]=0$}
\end{cases}
\end{equation}

Where 2$\Delta$ denotes how many positions can change in neighboring vectors at most. In Bloom filter encoding, $\Delta$ is equal to $k$ the size of the hash functions domain. $B^{'}$ represents the perturbed version of $B$.

\begin{figure}
\begin{center}
%\centering
\includegraphics[scale=0.60]{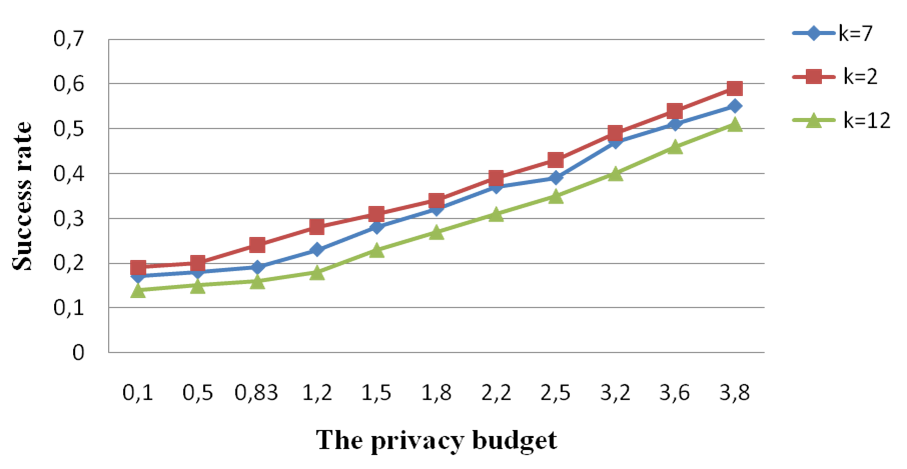}
\caption{Success rate of the plausible deniability attack 
\\ over one preference by a Basic Adversary, the preference dataset}
\label{one}
\end{center}
\end{figure}

\textbf{Experimental results for quantifying the chance of $BA$ of winning the game.}
The success rate of $BA$ for winning the game can directly be derived from the probability of Equation (\ref{dec}) as the probability of getting B[i]=1. This probability can be experimentally measured on our preference dataset, according to $\epsilon$ and $k$ values and gives the results depicted in Figure \ref{one}. As expected, the lower $\epsilon$ and the higher $k$, the lower the probability for winning the game is. For $\epsilon$ ranging from $[0.1,0.85]$ value, the probability that the adversary wins the game is in average below $0.22$. Higher the $k$ values, more difficult it is for $BA$ to win the game as the number of hash collisions increases, thus leading to higher wrong guess. 

As a conclusion, the success rate of $BA$ can be low according to the selected $\epsilon$ and $k$ values. As such, a suitable trade-off utility vs privacy, as detailed in Section \ref{sec:Trade-off}, needs to be found. 
%for $\epsilon$ ranging from $[0.1,0.85]$ value, the probability that the adversary wins the game is in average below $0.22$ and for higher $k$ values, the whole system becomes more resistant to the attack.

%Accordingly, the results of this game depend on the selected values of $\epsilon$ and $k$. Precisely, the lower and the higher $\epsilon$ and $k$, respectively are, the lower the probability to win the game is.
%To summarize, we can say that our system might be robust against one preference guessing attack in case of very low epsilon values and high $k$ values.

\subsection{Plausible Deniability over Multiple Preferences} 
\begin{figure}
\begin{center}
%\centering
\text{\ref{matr}.a [$\epsilon=0.10$]}
{\includegraphics[scale=0.60]{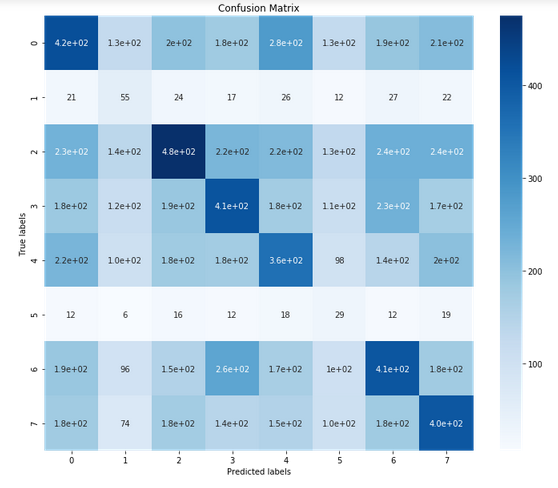}}
\text{\ref{matr}.b [$\epsilon=1.2$]}
{\includegraphics[scale=0.60]{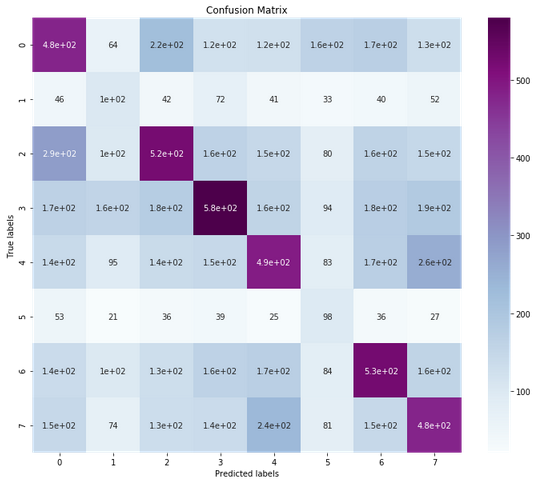}}
\text{\ref{matr}.c [$\epsilon=2.4$]}
{\includegraphics[scale=0.60]{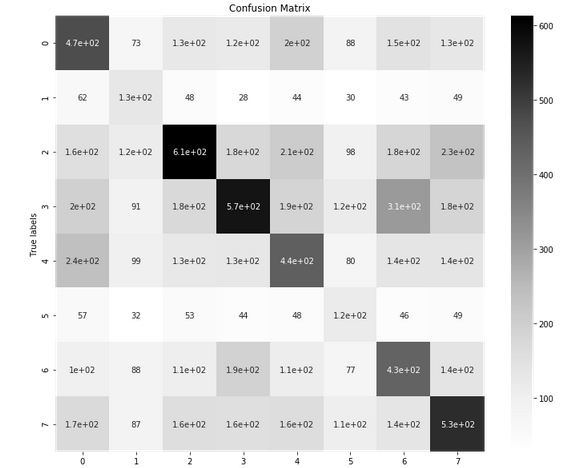}}
\caption{Decoding ability of an adversary for various privacy budget on Music Category. Test set=$10.000$ samples}
\label{matr}
\end{center}
\end{figure}

The attack is managed by an advanced adversary $AA$ as presented in Subsection \ref{sec:threat}. This section provides experimental results, and an in-depth discussion, about the $\epsilon$ impact on the success rate of the adversary. The experiment is conducted over 10.000 samples and 20.000 samples given to $AA$ for training.
Results are provided in two figures, Figure \ref{matr} for the confusion matrices, and Table \ref{atta_report} for the classification result statistics which enables to evaluate the decoding ability of the adversary. Table \ref{rate} gives the success rate of an $AA$ adversary to win the game.

%We provide an experimental proof for this game. Therefore, we simulated the experience while perturbing the attributes using various privacy budgets.
%We note that the tests were conducted over 10.000 samples and that the adversary was given 20.000 samples for training.

Figure \ref{matr}.a shows that for low privacy budget ($\epsilon = 0.1$), the adversary has mistaken the majority of the classes as the values outside the diagonal are relatively high. His overall precision and recall are below 29 $\%$ as demonstrated by the classification report in Table \ref{atta_report}.

An increase of the privacy budget from $\epsilon=0.1$ to $\epsilon=1.2$, resp. $\epsilon=2.4$, improves the classification ability of the adversary. Indeed, the precision and recall reach only 31 $\%$, resp. 49 $\%$ in average as depicted in Table \ref{atta_report}. Thus the success rate for winning the game by $AA$ $\epsilon=1.2$, resp. $\epsilon=2.4$, is equal to 33 $\%$, resp. 52 $\%$ (see Table \ref{rate}). One can also notice that for $\epsilon=2.4$, values inside the diagonal of the confusion matrix in Figure $\ref{matr}$.c are higher than for matrices in Figures \ref{matr}.a and \ref{matr}.b.

\begin{table*}
\begin{center}
\caption{The classification reports for different privacy budgets. Training set=$20.000$, test set=$10.000$ samples}
\resizebox{\textwidth}{!}{
\begin{tabular}{c|c|c|c|c|c|c|c|c|c|c}
\hline
 Metrics & \multicolumn{3}{c|}{ Precision } & \multicolumn{3}{c|}{Recall} & \multicolumn{3}{c|}{F1-score} & Support \\ \hline

& $\epsilon = 0.10$ & $\epsilon = 1.2 $ & $\epsilon = 2.4$ & $\epsilon = 0.10$ & $\epsilon = 1.2 $ & $\epsilon = 2.4$ & $\epsilon = 0.10$ & $\epsilon = 1.2 $ & $\epsilon = 2.4$ & $\epsilon = 0.10$, $1.2$, $2.4$ \\ \hline \hline
Classical & 0.24 & 0.32 & 0.51 & 0.29 & 0.33 & 0.54 & 0.26 & 0.33 & 0.52 & 1460  \\ \hline
Country & 0.27 & 0.24 & 0.45 & 0.08 & 0.14 & 0.31 & 0.12 & 0.18 & 0.37 & 716 \\ \hline
Electro & 0.25 & 0.32 & 0.51 & 0.34 & 0.37 & 0.57 & 0.29 & 0.35 & 0.54 & 1410  \\ \hline
Jazz & 0.26 & 0.34 & 0.55 & 0.29 & 0.41 & 0.50 & 0.27 & 0.37 & 0.52 & 1424 \\ \hline
Pop & 0.24 & 0.32 & 0.51 & 0.26 & 0.35 & 0.56 & 0.25 & 0.34 & 0.53 & 1396  \\ \hline
Rap & 0.23 & 0.29 & 0.43 & 0.04 & 0.14 & 0.32 & 0.07 & 0.19 & 0.37 & 717  \\ \hline
Rock & 0.26 & 0.36 & 0.49 & 0.28 & 0.37 & 0.52 & 0.27 & 0.36 & 0.50 & 1432 \\ \hline
Techno & 0.29 & 0.33 & 0.53 & 0.28 & 0.33 & 0.56 & 0.28 & 0.33 & 0.55 & 1445 \\ \hline

  \end{tabular}}
 \label{atta_report}
\end{center}
\end{table*}

%Only when we used a privacy budget that equals to $2.4$, the  decoding ability of the adversary reached 52 $\%$ like demonstrated in figure \ref{atta_report}.b. Also, the values inside the diagonal of the confusion matrix in figure $\ref{matr}$.c are higher when compared with the two other matrices (Figures \ref{matr}.a and \ref{matr}.b).
 
%In overall, the adversary has a low probability of winning the game as described in Table \ref{rate}.

\begin{table}
%\captionsetup{justification=centering}
\begin{center}
\caption{Success rate of the attack} 
%\centering
   \begin{tabular}{|c|c|}
   \hline
	Privacy budget & Success rate\\ \hline
	0.1 &29$\%$ \\
	\hline
	1.2&33 $\%$ \\
	\hline
	2.4 &52 $\%$ \\
	\hline
  \end{tabular}
\label{rate}
\end{center}
\end{table}

\subsection{Averaging Attack}
As described in Section \ref{sec:threat}, the adversary is provided with a number of perturbed Bloom filters corresponding to the same set of $C$'s preferences. The adversary is only able to compute the first PRR output value $B^{'}$ which is the first-level perturbed Bloom filter. As the same $B^{'}$ value is used from one session to another (cf. Equation \ref{IRR}), the adversary is unable to find out the original Bloom filter $B$. As such, he is not able to win the averaging attack game. 

\section{Privacy vs Utility Trade-off }
\label{sec:Trade-off}
This analysis measures the utility in terms of classification accuracy, which is defined as the ability of the recommender to perform correct classification of users, despite the perturbation scheme. Privacy is measured as the success rate of an advanced adversary for decoding the preferences. 

\begin{figure}%[H]
\begin{center}
%\centering
\text{\ref{trad}.a [Flight dataset]}
{\includegraphics[scale=0.35]{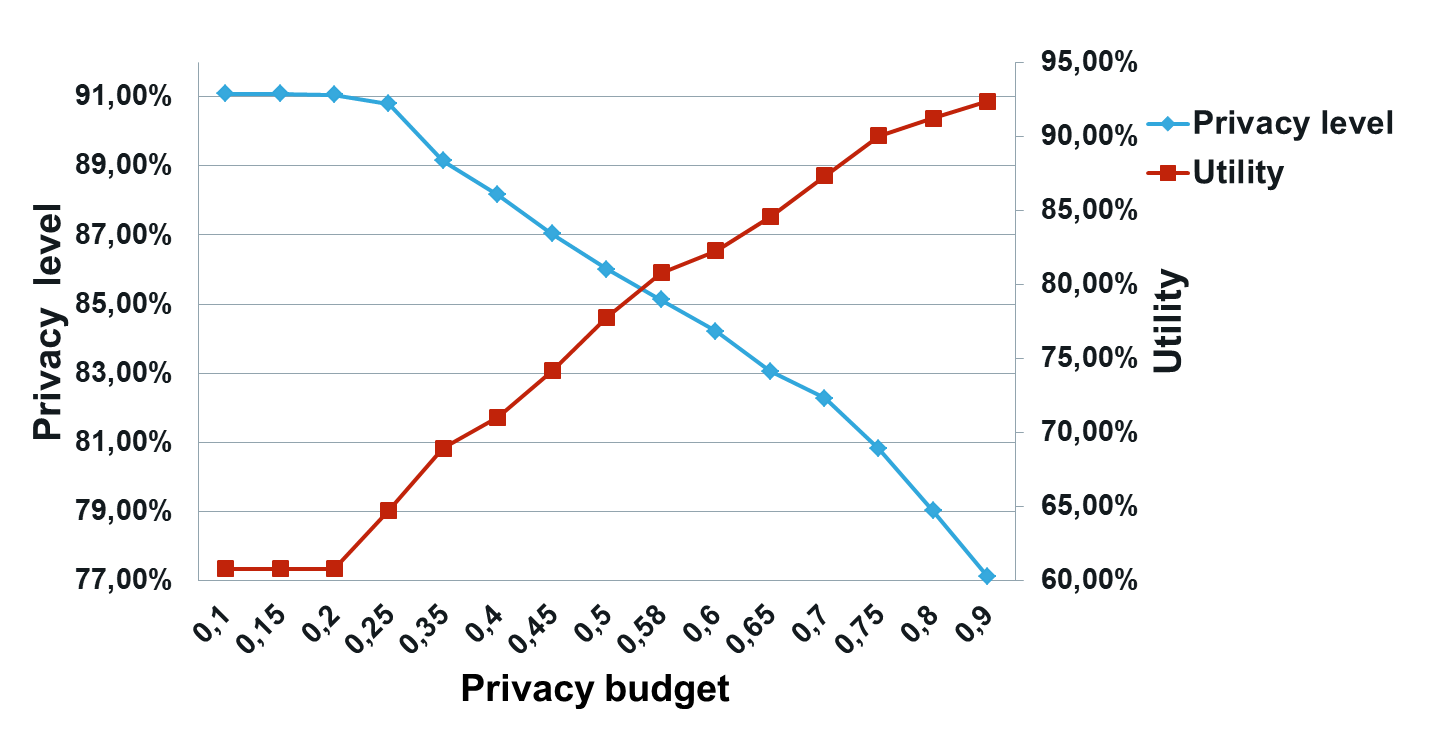}}
\text{\ref{trad}.b [Preference dataset]}
{\includegraphics[scale=0.40]{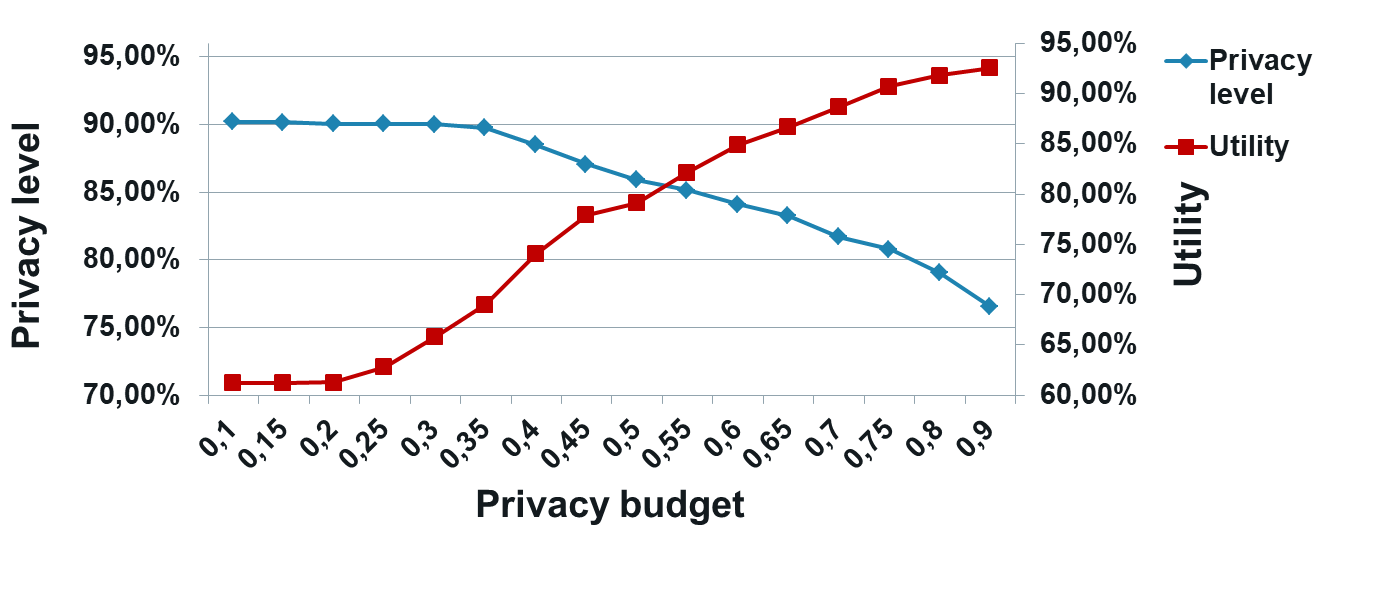}}
\caption{Privacy vs Utility plots for $90.000$ profiles, $k=15$, $K=5$}
\label{trad}
\end{center}
\end{figure}

The experimental setup considers $k=15$, 90.000 different profiles, and $K=5$ clusters. 
Figures \ref{trad}.a and \ref{trad}.b illustrate the achieved trade-off for both datasets with specific parameters $k=15$ and $K=5$. As expected, the privacy is a decreasing function of the privacy budget and the utility is a rising function of the privacy budget. There is a privacy vs utility trade-off (curves intersection) happening for $84 \%$ of privacy level and $80 \%$ of utility for both datasets for a privacy budget equal to $0.58$. This point is a kind of optimum when both utility and privacy are equally important.

%For all the values of the privacy budget that are lower than 0.58, the system achieves high privacy level at the expense of low utility. Conversely, privacy budget values that are higher than $0.58$ result in a important data utility level at the cost of an unreliable level of privacy.
%We note that the achieved trade-off may vary depending on the selected parameters of hash function, the profile size and the perturbation level desired by the client.

\section{Conclusion}
\label{sec:Conclusion}
This paper proposes a privacy-preserving recommandation system, which lets the user decide on the amount of preference data he wants to communicate to a recommender, and the amount of LDP noise he wants to introduce into his data. Thus the user can decide how much privacy vs user experience he wants to keep. Through our specific experiment conducted over two datasets, our solution exhibits good performances in terms of privacy and utility, i.e. a 90$\%$ clustering success rate, resp. 80.3$\%$ for a value of LDP $\epsilon = 0.8$, resp. $\epsilon = 2$, 
and it shows that a privacy vs utility trade-off can be found for $\epsilon=0.58$, with $84 \%$ privacy level and $80 \%$ utility. 

%Recommender systems are of paramount importance as they provide users with personalized content and contribute to be a major revenue driver for service providers. However, they pose a threat to users' privacy. To overcome the issue of massive users data collection, this paper proposes a system where users keep control over their data and decide how many data are communicated to the recommenders by introducing an LDP perturbation and get recommendations able to perturb their preferences using the local differential privacy notion. 

%Our approach includes two perturbation levels to boost the privacy at the user side. As for the service provider, in order to maintain a reliable level of utility, we designed a machine learning algorithm. The security of our solution is demonstrated while reference to two security games. 
%Clearly our solution exhibits performances in terms of privacy and utility when compared with other systems. Particularly, we presented the resistance against two security attacks.

\section{Acknowledgements}

This paper is supported in part by Institut Mines-Telecom chair VP-IP for Values and Policies of Personal Information and in part by the European Union's Horizon 2020 research and innovation program under grant agreement No 830892, SPARTA project. Authors are also thankful to Qwant for providing a dataset of cleaned traces which helped to achieve results closer to the ground.

%\section*{References}

\bibliographystyle{plain}
\bibliography{bibliography}

\end{document}